\documentclass[12pt]{article}

\advance \textheight by \topskip
\makeatletter
\@addtoreset{equation}{section}
 
\makeatother

\def\I{\hbox{1\hskip -3.2pt I}}

\def\OA{\mathfrak{OA}}
\def\gr{\mathop{\sf gr}}
\def\N{\mathbb N}
\def\R{\mathbb R}
\def\C{\mathbb C}
\def\Z{\mathbb Z}
\def\J{\mathcal J}

\usepackage{amsmath,amssymb}
\begin{document}

\title{
\begin{flushright}
{\small USACH-FM-01/11}\\[1.0cm]
\end{flushright}
{\bf  Nonlinear holomorphic supersymmetry, Dolan-Grady
relations and Onsager algebra
}}

\author{{\sf Sergey M. Klishevich${}^{a,b}$}\thanks{
E-mail: sklishev@lauca.usach.cl}
{\sf\ and Mikhail S. Plyushchay${}^{a,b}$}\thanks{
E-mail: mplyushc@lauca.usach.cl}
\\
{\small {\it ${}^a$Departamento de F\'{\i}sica,
Universidad de Santiago de Chile,
Casilla 307, Santiago 2, Chile}}\\
{\small {\it ${}^b$Institute for High Energy Physics,
Protvino, Russia}}}
\date{}

\maketitle

\vskip-1.0cm

\begin{abstract}
Recently, it was  noticed by us that the nonlinear
holomorphic supersymmetry of order $n\in\N,\,n>1$, 
($n$-HSUSY)
has an algebraic origin. We show that the Onsager algebra
underlies $n$-HSUSY and investigate the structure of the
former in the context of the latter. A new infinite set of
mutually commuting charges is found which, unlike those from
the Dolan-Grady set, include the terms quadratic in the
Onsager algebra generators. This allows us to find the
general form of the superalgebra of $n$-HSUSY and  fix it
explicitly for the cases of $n=2,3,4,5,6$. The similar
results are obtained for a new, contracted form of the
Onsager algebra generated via the contracted Dolan-Grady
relations. As an application, the algebraic structure of the
known 1D and 2D systems with $n$-HSUSY is clarified and a
generalization of the construction to the case of nonlinear
pseudo-supersymmetry is proposed. Such a generalization is
discussed in application to some integrable spin models and
with its help we obtain a family of quasi-exactly solvable
systems appearing in the $PT$-symmetric quantum mechanics.
\end{abstract}
\newpage

\section{Introduction}
Recently, there appeared a significant interest
in supersymmetry characterized by a
nonlinear superalgebra \cite{andrianov,susy-pb,
susy-pf,d1,Nfold,d2,dorey}.
It is a natural generalization of the usual supersymmetry
given by a linear superalgebra
\cite{witten,cooper}.
The nonlinear supersymmetry
emerges, in particular,
in the context of quantization
of the
one-dimensional mechanical systems possessing
supersymmetry of a general (linear or nonlinear) form
\cite{d1}.
It was proved \cite{d1} that
any classical 1D
supersymmetric system is symplectomorphic
to a supersymmetric system
of a canonical form in which
the supercharges are (anti)holomorphic
monomials of oscillator-like even and odd variables.
The anomaly-free quantization
of a canonical system with (anti)holomorphic supercharges
is possible only for the
oscillator-like bosonic operators of the special
restricted form.
As a result \cite{d1}, the anomaly-free quantization
leads to appearance of the
family of the 1D quantum systems with
nonlinear holomorphic supersymmetry of order
$n\in \N$, $n>1$, ($n$-HSUSY)
which turns out to be directly related to
the so called quasi-exactly solvable systems
\cite{zaslavskii,turbiner,shifman,ushv,olv,dunne}.
The nonlinear holomorphic supersymmetry of
the 1D quantum systems \cite{d1}
was generalized subsequently by us to the 2D
case represented by the system of
a charged particle on the plane
with magnetic field \cite{d2}.
It was found that the main difference of the 2D systems
with
holomorphic supersymmetry from the 1D case \cite{d1}
consists in appearance of an even central charge entering
non-trivially into the nonlinear superalgebra.
It was also observed \cite{d2} that the reduction of
the 2D systems with holomorphic supersymmetry
can produce nonlinear supersymmetry of a non-holomorphic
form \cite{Nfold,aoyama} related to
some families of quasi-exactly solvable systems.

In Ref. \cite{d2} we noted that the Hamiltonian of a
supersymmetric system and corresponding supercharges can be
realised in terms of two mutually conjugate operators
$Z$ and $\bar{Z}$
satisfying some nonlinear relations
(see Eq. (\ref{int}) below).
Such a formulation does not depend on the
choice of a particular representation for $Z$ and $\bar{Z}$.
This means that  $n$-HSUSY is based on
a universal algebraic foundation
\cite{d2,shrev}.
{\it The present paper is
devoted to the detailed investigation of the universal
algebraic
construction underlying the nonlinear holomorphic
supersymmetry}.

The paper is organized as follows.
In Section 2 we describe shortly the structure
of $n$-HSUSY
revealed in Ref. \cite{d2} and
discuss the nonlinear
algebraic relations underlying it.
The nonlinear relations turn out to be
a peculiar form of the
Dolan-Grady relations which appeared
earlier in the context of integrable models.
Two different cases of the
nonlinear relations have to be distinguished.
More simple case generating the contracted
Onsager algebra is investigated in Section 3. To our
knowledge, such a form
of the Onsager algebra was not discussed earlier
in literature. We find an infinite set
of the mutually commuting operators being quadratic
in generators of the contracted Onsager algebra.
It is this set of quadratic operators
that is associated with the
Hamiltonian and central elements of the nonlinear
superalgebra. The obtained results are applied
for finding the exact form of the nonlinear
superalgebra corresponding to the case
of $n$-HSUSY associated with
the contracted Onsager algebra.
In Section 4 we investigate in the same line
the non-contracted case of the Onsager algebra
and associated $n$-HSUSY.
In Section 5 we discuss the 1D and 2D
systems with nonlinear holomorphic supersymmetry
within the general algebraic framework
developed in the previous two sections.
A minimal modification of the general
algebraic construction allows us
to obtain the nonlinear pseudo-supersymmetry
which can be considered as a natural generalization
of the linear pseudo-supersymmetry \cite{phsym}.
In this way in section 6
we treat some integrable spin models
and obtain a family of
quasi-exactly solvable systems
appearing in the context of the $PT$-symmetric quantum
mechanics \cite{bb}.
In section 7 the brief summary of the obtained results is
presented.

\section{The structure of $n$-HSUSY
and Onsager algebra}

The holomorphic supersymmetry is constructed in terms of the
pairs of even, $Z$, $\bar Z=Z^\dag$, and odd, $\theta^+$,
$\theta^-=(\theta^+)^\dag$, operators. The odd operators
commute with the even ones and satisfy the
anticommutation relations: $\{\theta^-,\:\theta^+\}=1$,
$(\theta^\pm)^2=0$.
If the even operators
obey the nonlinear relations
\begin{align}\label{int}
 \left[Z,\left[Z,\left[Z,\bar Z\right]\right]\right]&=
 \omega^2\left[Z,\bar Z\right],&
 \left[\bar Z,\left[\bar Z,\left[Z,\bar Z\right]\right]
 \right]&= \bar\omega^2\left[Z,\bar Z\right]
\end{align}
with $\omega\in\mathbb C$, $\bar\omega=\omega^*$, then
for the Hamiltonian
of the form
\begin{equation}\label{Halg}
 H_n=\frac 12\left\{\bar Z,\:Z\right\}+
 \frac n2\left[Z,\:\bar Z\right]\cdot
 \left[\theta^+,\:\theta^-\right]
\end{equation}
the odd nilpotent operators defined by the recurrent
relations
\begin{align}
 Q_{n+2}&=\left(Z^2-\left(\tfrac{n+1}2\right)^2
 \omega^2\right)Q_n,&
 Q_0&=\theta^+,&Q_1&=Z\theta^+,
 \notag\\[-2.8mm]\label{Qalg}\\[-2.8mm]\notag
 \bar Q_{n+2}&=\left(\bar Z^2-\left(\tfrac{n+1}2\right)^2
 \bar\omega^2\right)\bar Q_n,&
 \bar Q_0&=\theta^-,&\bar Q_1&=\bar Z\theta^-,
\end{align}
are the integrals of motion,
$[Q_n,H_n]=[\bar Q_n,H_n]=0$,
being the supercharges
of the system (\ref{Halg}) \cite{d2}.
In what follows, we will refer to
the system with the Hamiltonian $H_n$ and
the supercharges
$Q_n$, $\bar Q_n$ defined by Eqs. (\ref{int})--(\ref{Qalg})
as to the nonlinear
holomorphic supersymmetric system.

Relations (\ref{int}) can be treated as
``integrability conditions'' of $n$-HSUSY.
For the linear case $n=1$,
unlike the nonlinear one, $n>1$,
the
restrictions (\ref{int}) on the operators $Z$, $\bar Z$
are not necessary since the supersymmetry,
\begin{align*}
[Q_1,H_1]&=0,&[\bar Q_1,H_1]&=0,&
\{Q_1,\bar Q_1\}&=H_1,&
Q_1^2&=\bar Q_1^2=0,
\end{align*}
appears just due to the
special form of the Hamiltonian.
Hence, the
$n=1$ supersymmetry has a clear algebraic origin different
from that of the nonlinear case $n>1$.
Let us note also
that such a linear supersymmetry is always
holomorphic by the construction. In this sense, the
nonlinear holomorphic supersymmetry is a natural algebraic
generalization of the linear supersymmetry.

In the 1D case \cite{d1}, $n$-HSUSY
is obtained via the realization
$Z=\frac{d}{dx}+W(x)$,
$\bar Z=-\frac d{dx}+W(x)$, where $W(x)$ is a real function
(superpotential) defined by the relations (\ref{int}).
In this representation the supercharges and the Hamiltonian
of
one-dimensional system with $n$-HSUSY
form a polynomial
superalgebra, $\{Q_n,\bar Q_n\}=P_n(H_n)$,
where $P_n$ is the $n$th degree polynomial.
The supercharges, the
Hamiltonian and the even central charge
appearing in the 2D plane case \cite{d2}
also form a
nonlinear superalgebra of the polynomial type,
but there the anticommutator of the supercharges
produces a polynomial in both even integrals of motion.
As we shall see,
generally a system with $n$-HSUSY
admits an infinite number of the central charges.
They enter nontrivially into the superalgebra
via the anticommutator of the supercharges
being a polynomial in the Hamiltonian as well as
in the central charges.
However, depending on the concrete representation of the
operators
$Z$ and $\bar Z$, some of the central charges can
turn into zero,
or can be representable in terms of other central charges.
Note that
the polynomiality of the superalgebra
makes $n$-HSUSY to be
similar to the
Yangian and finite W-algebras \cite{yangian1,yangian2,Walg}.

The amazing fact is that the ``integrability
conditions'' (\ref{int}) are, in fact, the well-known
Dolan-Grady relations \cite{dg}. For the first time such
relations appeared in the context of integrable models
\cite{dg}. In Ref. \cite{dg} it was proved that relations of
the form (\ref{int}) allow ones to construct the
infinite number
of commuting operators. It is such a set of mutually
commuting operators that relates naturally a
construction to integrable models.
In the most elegant way the
construction can be represented in terms of the Onsager
algebra ($\OA$), which is recursively generated by means of
the Dolan-Grady relations \cite{dg}. The Onsager algebra
\cite{onsager} is spanned by the generators $A_n$
and $G_n$
having the commutation relations
\begin{align}\label{OA0}
 \left[A_m,\:A_n\right]&=4G_{m-n},&
 \left[G_m,\:A_n\right]&=2A_{n+m}-2A_{n-m},&
 \left[G_m,\:G_n\right]&=0,
\end{align}
where $m,n\in\Z$.
In fact, the operators $A_0$ and $A_1$ are
the generating elements of $\OA$ obeying the Dolan-Grady
relations while all the other operators are defined
recursively by the relations (\ref{OA0}) \cite{dav1,dav2}.
Actually, we have
the correspondence $A_0\sim\bar Z$, $A_1\sim Z$.
However, unlike the present case (\ref{int}),
the operators $A_0$, $A_1$ are usually
supposed to be Hermitian. The conventional commutation
relations (\ref{OA0}) correspond to the choice
$\omega^2=16$, which for $\omega\neq0$
can always be achieved via
the appropriate
rescaling of the operators.
Though the generating elements of
$\OA$, $A_0$ and $A_1$,
are usually supposed to be Hermitian,
they simultaneously are related by some
kind of duality~\cite{dg}. This generates an involution on
the whole Onsager algebra to be additional to the Hermitian
conjugation.

The Onsager algebra finds numerous applications in
studying integrable models \cite{dg,dav1,dav2,gehlen}.
Specifically, the infinite set
of the mutually
commuting operators constructed by Dolan and Grady can be
represented as
\begin{equation}\label{lin}
 2J_m=A_m+A_{-m}+\lambda\left(A_{m+1}+A_{1-m}\right),
\end{equation}
where $\lambda\in\R$.
The commutativity of the operators $J_m$
can easily be verified within the framework of the algebra
(\ref{OA0}) (cf. with the
original proof of Ref. \cite{dg}).
Then, treating the operator
$J_0=A_0+\lambda A_1$ as a Hamiltonian,
one obtains,
generally, an integrable system with infinite number of
conserved commuting charges. In various models, such as the
Baxter eight-vertex
\cite{baxter1,baxter2,baxter3}, the
two-dimensional Ising \cite{onsager} and $Z_N$ spin models
\cite{gehlen}, there appear the charges of the form
(\ref{lin})
being equivalent to those
which result
from the transfer matrix formulation and exact
quantum integrability of the system. The power of the
algebraic formulation (\ref{OA0}), (\ref{lin}) is that it
does not refer to the number of dimensions or to
the nature of
the space-time manifold which can be lattice, continuum or
loop
space.
However, we should note that till now the application of the
construction to dimensions $D>1$ is an open
problem.

As we will see, the set (\ref{lin}) is not a
unique set of mutually
commuting operators that can be constructed in terms
of generators of the Onsager algebra.
It turns out that
there exists also a
set of commuting operators to be quadratic in the
generators of $\OA$. Such commuting operators naturally
appear in the scheme of the nonlinear holomorphic
supersymmetry.

\section{Contracted Onsager algebra}

Before discussing the general case of relations (\ref{int})
with $\omega\neq0$ and the
corresponding Onsager algebra, let us
consider a more
simple reduced case with
$\omega=0$. Such a construction
is also of a
significant physical interest. For example,
the family of quasi-exactly solvable
systems with the $x^6$-potential
is produced by a reduction
of the plane system with $n$-HSUSY
in which $\omega=0$ \cite{d2}.
It is worth also noting that
$n$-HSUSY admits
a generalization to
the systems with nontrivial 2D Riemannian geometry only
in this
case \cite{sphere}.

The operators $Z_0\equiv Z$ and $\bar Z_0\equiv \bar Z$
together with the contracted ($\omega=\bar\omega=0$) Dolan-
Grady relations
(\ref{int}) recursively generate the following
infinite-dimensional algebra:
\begin{align}
 \left[Z_m,\:\bar Z_n\right]&=B_{m+n+1},&
 \left[Z_n,\:B_m\right]&=Z_{m+n},&
 \left[B_m,\:\bar Z_n\right]&=\bar Z_{m+n},
 \notag\\[-2.5mm]\label{A}\\[-2.5mm]\notag
 \left[Z_m,\:Z_n\right]&=0,&
 \left[\bar Z_m,\:\bar Z_n\right]&=0,&
 \left[B_m,\:B_n\right]&=0,
\end{align}
where $m,n\in\Z_+$ and $B_0=0$ is implied.
We denote this contracted Onsager algebra as
$\mathfrak{COA}$
and name operators $Z_0$, $\bar Z_0$ the generating
elements of the algebra~(\ref{A}).

Algebra $\mathfrak{COA}$
possesses the involution
defined by the relations
\begin{equation}\label{invol}
 \bar Z_m=Z_m^\dag,\hskip 2cm B_m^\dag=B_m.
\end{equation}
Besides, the natural grading
is induced
in it by putting
$\gr Z_0=\gr\bar Z_0=1$:
\begin{align}\label{gr}
 \gr Z_m&=\gr\bar Z_m=2m+1,&\gr B_m&=2m.
\end{align}
This grading turns out to be
a useful tool in
practical calculations,
e.g., in computing nonlinear
superalgebra.

One can verify that the operators $Z_k$, $\bar Z_k$ with
$k\in\N$ also obey the contracted Dolan-Grady
relations
\begin{align}\label{int0}
 \left[Z_k,\:\left[Z_k,\:\left[Z_k,\:\bar Z_k\right]\right]
 \right]&=0,&
 \left[\bar Z_k,\:\left[\bar Z_k,\:\left[Z_k,\:\bar Z_k
 \right]\right]\right]&=0.
\end{align}
This means that one can consider
the operators $Z_k$, $\bar Z_k$ as a new pair
of generating elements,
$Z_0^{(k)}$, $\bar Z_0^{(k)}$, which induce an
infinite-dimensional subalgebra
$\mathfrak {COA}_k\subset\mathfrak {COA}$. The
inclusion $\mathfrak {COA}_k\to\mathfrak {COA}$ is given by
\begin{align}\label{incl}
 Z_m^{(k)}&=Z_{2km+k+m},&
 \bar Z_m^{(k)}&=\bar Z_{2km+k+m},&
 B_m^{(k)}&=B_{m(2k+1)}.
\end{align}
{}From (\ref{int0}) it follows that
for any
$k\in\mathbb N$
the algebra
$\mathfrak {COA}_k$ itself is a contracted Onsager
algebra, i.e. we have the natural isomorphism,
$\mathfrak {COA}_k\simeq\mathfrak {COA}$. Obviously, any
algebra $\mathfrak {COA}_k$ also has an
infinite number of isomorphic subalgebras and so on. This
property resembles fractals in the sense that the algebra
consists of infinite subsets with the very algebraic
properties of the initial set and at any level of
embedding we observe the same algebraic structure. Moreover,
this property leads to the following corollary.
First, we note that the operators (\ref{lin})
commute with each other in the case of the contracted
Onsager algebra as well (see the identification
(\ref{Zdef}) between the
generators $A_n$, $n\in\Z$,
and $Z_m$, $\bar Z_m$, $m\in\Z_+$, below).
Then, for any
given representation of the generating elements $A_0$,
$A_1$, besides the main set of the commuting charges
(\ref{lin}), in general, there exist additional
independent sets associated with the sequence of the
subalgebras%
\footnote{Since
$\mathfrak {COA}_m\simeq\mathfrak {COA}$ for any
$m\in\N$,
such a sequence is defined for every subalgebra
$\mathfrak{COA}_m$ as well, and so on.} $\mathfrak {COA}_1$,
$\mathfrak {COA}_2$, $\ldots$. If a representation of
the operators $A_0$, $A_1$ permits to generate the whole
infinite-dimensional algebra $\mathfrak {COA}$
with linearly independent generators,
then the number of such sets is infinite.

It is known that the Onsager algebra (\ref{OA0})
can be regarded as a fixed subalgebra of the $sl_2$-loop
algebra, $C[t,\:t^{-1}]\otimes sl_2$ \cite{roan}.
Such a representation
facilitates a systematic study of the algebraic structure of
the Onsager algebra \cite{roan}. The contracted algebra
$\mathfrak {COA}$
can also be represented as a loop extension of the
$sl_2$ algebra:
\begin{align}
 Z_m&=\frac 1{\sqrt 2}t^{2m+1}L_-,&
 \bar Z_m&=\frac 1{\sqrt 2}t^{2m+1}L_+,&
 B_m&=\frac 12t^{2m}L_0,\label{sl2}
\end{align}
where $m\in\Z_+$ and
operators $L_\pm$, $L_0$ are the
generators of the
$sl_2$ algebra,
$[L_-,L_+]=L_0$, $[L_0,L_\pm]=\pm 2L_\pm$.
Note here that the degrees of $t$
in generators of $\mathfrak {COA}$ coincide with
the corresponding grading
numbers~(\ref{gr}).
The
representation (\ref{sl2})
implies that the algebra $\mathfrak {COA}$
is a
subalgebra of the loop algebra $C[t]\otimes sl_2$.

Let us consider the following infinite set of operators:
\begin{equation}\label{jmn}
 J_n^m=\frac 12\sum_{p=1}^m\left(
 \left\{\bar Z_{p-1},\:Z_{m-p}\right\}-B_pB_{m-p}\right)
 -\frac n2B_m,
\end{equation}
where $m\in\N$ and $n$ is an arbitrary
real number.
Using the commutation relations (\ref{A}),
one can demonstrate directly
the commutativity of all these
operators with the same value of the parameter
$n$,
\begin{equation}\label{com}
 \left[J^p_n\,,\:J^m_n\right]=0.
\end{equation}
So, in addition to
the set of commuting operators (\ref{lin}) linear in
generators of $\mathfrak {COA}$, there exits the infinite
set
of the
commuting operators (\ref{jmn}) quadratic
in generators of $\mathfrak {COA}$.

Now, we pass over to the construction of
$n$-HSUSY and
introduce the operators
\begin{equation}\label{sj}
 2\mathcal J^m_n=J^m_n\theta^-\theta^+
 +J^m_{-n}\theta^+\theta^-,
\end{equation}
which give a supersymmetric extension of
(\ref{jmn}).
Assuming $n\in\N$
in Eq. (\ref{jmn}), one finds
that the operator
$\J^1_n$ coincides with the $n$-supersymmetric Hamiltonian
(\ref{Halg}).
Then the set of even mutually commuting operators
(\ref{sj}) can be extended by the
odd nilpotent operators
\begin{equation}\label{q0}
 Q_n=Z_0^n\theta^+,\hspace{2cm}
 \bar Q_n=\bar Z_0^n\theta^-.
\end{equation}
Using the algebra (\ref{A}) and the
relations
\begin{align*}
 \left[Z_0^n,\:\bar Z_m\right]&=
 \frac{n(n-1)}2Z_{m+1}Z_0^{n-2}+nB_{m+1}Z_0^{n-1},&
 \left[Z_0^n,\:B_m\right]=nZ_mZ_0^{n-1}
\end{align*}
following from it,
one can demonstrate that the operators (\ref{sj})
commute with the odd ones,
${\left[Q_n,\:\mathcal J^m_n\right]=0}$. This means that the
operators (\ref{q0}) can be treated
as the supercharges
of the $n$-supersymmetric system
described by means of the Hamiltonian
$\mathcal J^1_n$, in which the set
$\mathcal J^k_n$, $k=2,3,\ldots$, plays
the role
of the even integrals of motion.
The corresponding nonlinear
superalgebra is polynomial and centrally extended.
For
$n=2,\:3,\:4,\:5,\:6$, the corresponding
anticommutators of the supercharges\footnote{For $n=1$, one
has the usual linear superalgebra.} are
\begin{align}
 \left\{Q_2,\:\bar Q_2\right\}&=\bigl(\mathcal J^1_2\bigr)^2
 +\frac 12\mathcal J^2_2,\notag\\\notag
 \left\{Q_3,\:\bar Q_3\right\}&=\bigl(\mathcal J^1_3\bigr)^3
 +2\mathcal J^1_3\mathcal J^2_3+\mathcal J^3_3,\\
 \left\{Q_4,\:\bar Q_4\right\}&=\bigl(\mathcal J^1_4\bigr)^4
 +5\bigl(\mathcal J^1_4\bigr)^2\mathcal J^2_4
 +6\mathcal J^1_4\mathcal J^3_4+\frac 94\bigl(\mathcal J^2
 _4\bigr)^2
 +\frac 92\mathcal J^4_4,
 \notag\\[-3mm]\notag\\[-3mm]
 \left\{Q_5,\:\bar Q_5\right\}&=\bigl({\cal J}^1_5\bigr)^5
 + 10\bigl({\cal J}^1_5\bigr)^3{\cal J}^2_5
 + 21\bigl({\cal J}^1_5\bigr)^2{\cal J}^3_5
 + 36{\cal J}^1_5{\cal J}^4_5
 + 16{\cal J}^1_5 \left({\cal J}^2_5\right)^2
 \label{rsalg}\\\notag&\hskip 5mm{}
 + 24{\cal J}^2_5{\cal J}^3_5 +36 {\cal J}^5_5,
 \\[-3mm]\notag\\[-3mm]\notag
 \left\{Q_6,\:\bar Q_6\right\}&=\left({\cal J}^1_6\right)^6
 + \frac{35}2\left({\cal J}^1_6\right)^4{\cal J}^2_6
 +56\left({\cal J}^1_6\right)^3{\cal J}^3_6
 +\left({\cal J}^1_6\right)^2\left(\frac{259}4
 \left({\cal J}^2_6\right)^2+162{\cal J}^4_6 \right)
 \notag\\&\hskip 5mm{}
 +{\cal J}^1_6\left(220{\cal J}^2_6{\cal J}^3_6
 +360{\cal J}^5_6\right)
 +\frac{225}8\left({\cal J}^2_6\right)^3
 +225{\cal J}^2_6{\cal J}^4_6
 +100\left({\cal J}^3_6\right)^2+450{\cal J}^6_6.
 \notag
\end{align}
{}From these explicit particular
relations one can naturally conjecture
that
the anticommutator of the corresponding supercharges is
proportional to a polynomial in the generators
${\cal J}_n^1$, ${\cal J}_n^2$, ..., ${\cal J}_n^n$ for
arbitrary $n$ as well.
Note that the grading (\ref{gr}) facilitates
calculation of the nonlinear superalgebra fixing
the general form of the polynomial up to numerical
coefficients.
However, we have not succeeded in fixing the corresponding
coefficients for the general case $n\in \N$.

\section{Onsager algebra}

Let us consider now the case of the
non-contracted ($\omega\ne 0$) Onsager algebra in
the context of $n$-HSUSY.
The nonlinear holomorphic supersymmetry
is constructed in terms of the
generators $Z$, $\bar Z$ obeying the relations (\ref{int}),
for which the
involution $Z^\dag=\bar Z$
is assumed. It is
clear that this operation defined on the generating elements
of $\OA$ induces the involution on the whole Onsager
algebra (\ref{OA0}). We treat this involution as a Hermitian
conjugation. To make this involution apparent, it is
convenient to adopt the notations
\begin{align}\label{Zdef}
 A_m&=2\sqrt 2 Z_{m-1},&
 A_{-n}&=-2\sqrt 2\bar Z_n,&
 G_{-m}&=2B_m,
\end{align}
where $m\in\mathbb N$, while $n\in\mathbb Z_+$. Then the
Onsager algebra can be rewritten in the form
\begin{align}
 \bigl[Z_n,\:B_m\bigr]&=Z_{n+m}-\alpha^{2m}Z_{n-m}
 +\alpha^{2n+1}\bar Z_{m-n-1},
 \notag\\\notag
 \bigl[B_m,\:\bar Z_n\bigr]&=\bar Z_{n+m}-
 \alpha^{2m}\bar Z_{n-m}+\alpha^{2n+1}Z_{m-n-1},
 \notag\\\notag
 \bigl[Z_n,\ Z_{n'}\bigr]&=
 \alpha^{2\,{\sf min}(n,n')+1}B_{n'-n},
 \notag\\[-3mm]\label{OA}\\[-3mm]\notag
 \bigl[\bar Z_n,\ \bar Z_{n'}\bigr]&=
 \alpha^{2\,{\sf min}(n,n')+1}B_{n-n'},
 \notag\\\notag
 \bigl[Z_n,\ \bar Z_{n'}\bigr]&=B_{n+n'+1},
 \notag\\\notag
 \bigl[B_m,\ B_{m'}\bigr]&=0,
\end{align}
where $n,n'\in\Z_+$,
$m,m'\in \N$, ${\sf min}(n,n')=n$ (or $n'$)
for $n\geq n'$ (for $n'>n$),
and we imply that the operators
$Z_k$, $\bar Z_k$ with $k<0$ and $B_0$ vanish in r.h.s.
of (\ref{OA}).
For the sake of simplicity here we
put $\omega^2=2\alpha$, $\alpha\in\R$, since we consider the
Onsager algebra over $\C$ and the parameter $\omega$ can
always be chosen in this form by
the appropriate rescaling of
the generators. Now one can make sure that the algebra $\OA$
(\ref{OA}) possesses the involution defined by the
relations (\ref{invol}).

Although representation
(\ref{OA}) is more complicated then the original form
(\ref{OA0}), besides more transparent structure
with respect to the conjugation (\ref{invol}),
it has two further advantages:
{\it a)} in notations (\ref{Zdef}) the contracted
algebra (\ref{A}) is reproduced easily in the limit
$\alpha\to 0$; {\it b)} these notations clarify
also the
property of self-similarity (``fractal'' structure)
of the Onsager algebra.
Indeed, like the generating elements $Z_0=Z$,
$\bar Z_0=\bar Z$,
the operators
$Z_m$, $\bar Z_m$, $m\in\N$, obey the Dolan-Grady
relations:
\begin{align}
 \notag
 \left[Z_m,\:\left[Z_m,\:\left[Z_m,\:\bar Z_m\right]\right]
 \right]&=2\alpha^{2m+1}\left[Z_m,\:\bar Z_m\right],
 \\[-2.5mm]\label{int1}\\[-2.5mm]
 \left[\bar Z_m,\:\left[\bar Z_m,\:\left[Z_m,\:\bar Z_m
 \right]\right]
 \right]&=2\alpha^{2m+1}\left[Z_m,\:\bar Z_m\right].
 \notag
\end{align}
This means that one can treat the operators $Z_m$,
$\bar Z_m$, $m\in\mathbb N$, as new generating elements,
$Z_0^{(m)}$, $\bar Z_0^{(m)}$, producing the
infinite-dimensional subalgebra $\OA_m\subset\mathfrak{OA}$.
The inclusion $\OA_m\to\OA$ is given by the same relations
(\ref{incl}) as for the contracted case. So,
$\OA$ also possesses the infinite set of isomorphic
subalgebras, $\mathfrak{OA}_m\simeq\mathfrak{OA}$,
$m\in\mathbb N$. As a consequence, additional sets
of mutually commuting operators are associated with every
subalgebra
$\OA_m$. If,
like in the contracted case,
the representation of the operators $Z_0$,
$\bar Z_0$ permits to generate the whole
infinite-dimensional algebra $\OA$
with linearly independent generators,
then the number of such sets is infinite.

By direct calculation one can verify that
as in the contracted case, the operators
(\ref{jmn}) form the infinite set of
mutually commuting charges for any value of the parameter
$n$ (not only integer) playing the role
of the coupling constant. The operator
$J_n^1$ can be treated as
the Hamiltonian of the system. Then, to
realize the nonlinear holomorphic supersymmetry we have to
restrict the parameter $n$ to integer values. The
supersymmetric extension of the
even charges is defined by Eq. (\ref{sj}).

By mathematical induction
one can prove that the odd
nilpotent operators defined by Eqs. (\ref{Qalg}) commute
with all the charges (\ref{sj}).
Let us sketch the proof. The
commutativity for $n=0$ and $n=1$ can be verified
by the direct calculation. Therefore, one have to prove
that
$[Q_{n+2},{\cal J}^m_{n+2}]=0$, $m\in\N$,
presuming that the relations
$[Q_n,{\cal J}^m_n]=0$ are valid.
The last equality
can be
represented as
\begin{align}\label{pr1}
 \left[{\cal Q}_n,J^m_n\right]-nB_m{\cal Q}_n=0,
\end{align}
where ${\cal Q}_n=\{Q_n,\:\theta^-\}$. Using (\ref{pr1}) and
the recursive relations (\ref{Qalg}) defining the
supercharges, the similar equality for $n\to n+2$ can be
represented in the form
\begin{align}
 \left[Z_0^2,J^m_{-n-2}\right]\left(Z_0^2-\alpha
 \frac{(n-1)^2}2
 \right)-\left(Z_0^2-\alpha \frac{(n+1)^2}2
 \right)\left[Z_0^2,J^m_{2-n}\right]=0.
\end{align}
The last identity can be verified directly,
and this proves that the odd nilpotent operators
(\ref{Qalg}) are the true supercharges
of the system with the
Hamiltonian ${\cal J}^1_n$.
It is clear that the operators
${\cal J}^m_n$, $m=2,3,\ldots$, are the
central charges of
the system. The corresponding superalgebra is polynomial and
centrally extended. For $n=2,\:3,\:4,\:5,\:6$
the anticommutators of the corresponding supercharges are
given by the following relations:
\begin{align}
 \notag
 \left\{Q_2,\:\bar Q_2\right\}&=\bigl(\mathcal J^1_2\bigr)^2
 +\tfrac 12\left(\mathcal J^2_2+\tfrac 12\alpha^2\right),\\
 \left\{Q_3,\:\bar Q_3\right\}&=\bigl(\mathcal J^1_3\bigr)^3
 +2\mathcal J^1_3\left(\mathcal J^2_3+\tfrac 12\alpha^2
 \right)+\mathcal J^3_3,
 \notag\\[-3mm]\notag\\[-3mm]\notag
 \left\{Q_4,\:\bar Q_4\right\}&=\bigl(\mathcal J^1_4\bigr)^4
 +5\bigl(\mathcal J^1_4\bigr)^2\left(\mathcal J^2_4
 +\tfrac 12\alpha^2\right)+6\mathcal J^1_4\mathcal J^3_4\\
 &\ \ \ +\tfrac 94\left(\mathcal J^2_4+\tfrac 12\alpha^2
 \right)^2+\tfrac 92\left(\mathcal J^4_4+\alpha^4\right),
 \notag\\[-3mm]\notag\\[-3mm]\notag
 \left\{Q_5,\:\bar Q_5\right\}&=\bigl({\cal J}^1_5\bigr)^5
 +10 \bigl({\cal J}^1_5\bigr)^3\left({\cal J}^2_5
 +\tfrac 12\alpha^2\right) + 21 \bigl({\cal J}^1_5\bigr)^2
 {\cal J}^3_5+36 {\cal J}^1_5
 \left({\cal J}^4_5+\alpha^4\right)
 \\&\ \ \ +16 {\cal J}^1_5 \left({\cal J}^2_5
 +\tfrac 12\alpha^2\right)^2 + 24 \left({\cal J}^2_5
 +\tfrac 12\alpha^2\right) {\cal J}^3_5
 +36 {\cal J}^5_5,
 \label{salg}\\[-3mm]\notag\\[-3mm]\notag
 \left\{Q_6,\:\bar Q_6\right\}&=\left({\cal J}^1_6\right)^6
 + \tfrac{35}2\left({\cal J}^1_6\right)^4
 \left({\cal J}^2_6+\tfrac 12\alpha^2\right)
 +56\left({\cal J}^1_6\right)^3{\cal J}^3_6
 \notag\\&\ \ \
 +\left({\cal J}^1_6\right)^2 \left(\tfrac{259}4
 \left( {\cal J}^2_6+\tfrac 12\alpha^2\right)^2
 +162 \left({\cal J}^4_6+\alpha^4\right) \right)
 \notag\\&\ \ \
 +{\cal J}^1_6\left(220\left({\cal J}^2_6
 +\tfrac 12\alpha^2\right){\cal J}^3_6+360{\cal J}^5_6
 \right) +\tfrac{225}8\left({\cal J}^2_6+\tfrac 12\alpha^2
 \right)^3\notag\\&\ \ \
 +225\left({\cal J}^2_6+\tfrac 12\alpha^2\right)
 \left({\cal J}^4_6+\alpha^4\right)
 +100\left({\cal J}^3_6\right)^2
 +450\left({\cal J}^6_6+\tfrac 32\alpha^6\right).
 \notag
\end{align}
It is worth noting that by the substitution
${\cal J}^{2k}_n+\frac{k}2\alpha^{2k}\to{\cal J}^{2k}_n$,
$k\in\N$, the relations (\ref{salg}) are reduced
exactly to the relations
(\ref{rsalg})
corresponding to the case
of the
contracted Onsager algebra.
One can naturally suppose that the same is valid for any $n$
as well. Presently we cannot explain this fact but
let us point out the analogy with the usual linear
supersymmetry. In the latter case the arbitrariness in the
energy shift is used to represent the superalgebra in the
conventional form (without constant term having a sense
of the central charge in the supercharges' anticommutator),
while in
the case (\ref{salg}) the shifts of the operators
${\cal J}^{2k}_n$ allow ones to represent the superalgebra
in the universal
form independent from the parameter $\alpha$.
It is necessary to note also that the calculation of the
superalgebra (\ref{salg}) has essentially facilitated
the derivation of the exact form of the even
central charges.

\section{1D and 2D $n$-HSUSY from the algebraic viewpoint}
As we have noted,
the nonlinear holomorphic supersymmetry was discussed for
the
first time in
Ref.~\cite{d1}. The holomorphic supersymmetry
arises naturally
in pseudo-classical systems possessing the supersymmetry of
the most general form. The Dolan-Grady relations appeared
then as
conditions of anomaly-free quantization of such classical
systems with {\it nonlinear} holomorphic supersymmetry. In
Ref.~\cite{d2} the construction was generalized to the case
of the quantum mechanics on a plane.

In the present section we
discuss these quantum systems with nonlinear holomorphic
supersymmetry within the algebraic framework developed above
proceeding from the different
explicit realizations of
the operators $Z$, $\bar Z$ obeying the Dolan-Grady
relations.

\subsection{1D $n$-HSUSY}
In one-dimensional quantum-mechanical systems
the nonlinear holomorphic supersymmetry
is generated by the linear differential operators
\begin{align}\label{zd1}
 Z_0&=\frac d{dx}+W(x),&\bar Z_0&={}-\frac d{dx}+W(x)
\end{align}
given in terms of the superpotential $W(x)$ being a real
function \cite{d1}.
The form of the superpotential is fixed by
the Dolan-Grady relations (\ref{int}),
and in the non-contracted case
($\omega^2=2\alpha\neq 0$)
the corresponding solution is
\begin{equation}\label{wd1}
 W(x)=w_+e^{\sqrt{2\alpha}x}+w_-e^{-\sqrt{2\alpha}x}+w_0,
\end{equation}
where $w_\pm$, $w_0\in\R$, are the parameters of the
system.
In the representation
(\ref{zd1}), (\ref{wd1}),
the generating elements give rise to the following
finite-dimensional algebra:
\begin{align}
 [Z_0,\:\bar Z_0]&=B_1,&
 [Z_0,\:B_1]&=G,&
 [Z_0,\:G]&=2\alpha B_1,
 \notag\\[-3mm]\label{ad1}\\[-3mm]\notag
 [B_1,\:G]&=0,&
 [B_1,\:\bar Z_0]&=G,&
 [G,\:\bar Z_0]&=2\alpha B_1,
\end{align}
with the generator $G$ satisfying,
by definition, the relation $G^\dag=G$.
Formally,
this algebra admits the two central
elements
\begin{align*}
 C&=2\alpha(Z_0+\bar Z_0)-G,&
 I&=G^2-2\alpha B_1^2,
\end{align*}
the second from which is the
quadratic Casimir operator of the algebra.
For any irreducible representation of the algebra
(\ref{ad1}) these operators have to be proportional to the
unit operator.
In representation
(\ref{zd1}), (\ref{wd1}), one finds that
$C=4\alpha w_0\cdot\I$ while $I=2^6\alpha^2w_+w_-\cdot\I$.
Therefore, this representation is irreducible:
for $\alpha>0$ the given
representation corresponds to the $iso(1,1)$
algebra,
whereas the $e(2)$ algebra
is produced for $\alpha<0$.
Indeed, in representation (\ref{zd1}), (\ref{wd1})
the corresponding linearly independent
nontrivial generators
can be represented
as
$\{i\frac d{dx}, e^{\pm x}\}$
($\alpha=\frac 12$),
or in the form
$\{i\frac d{dx}, e^{\pm ix}\}$
($\alpha=-\frac 12$).
The first set produces the Poincar\'e
algebra $iso(1,1)$, while the
second generates the Euclidean algebra $e(2)$.

It is interesting to note that though originally the
1D supersymmetric system is defined by the 4-parametric
superpotential (\ref{wd1}), the number of independent
continuous parameters is equal to two. Rescaling $x$,
one can reduce the parameter $\alpha$ to $\pm 1$,
whereas
a shift in $x$ results in rescaling the parameters
$w_\pm$. As a result,
the independent combinations of the parameters
characteristic to the system possessing
$n$-HSUSY are $w_0$ and $w_+w_-$.
These parameters define the values of the
central elements of the algebra (\ref{ad1}).

The relations (\ref{ad1}) give rise to
the Onsager algebra (\ref{OA}).
However,
in this case the higher generators
are linearly dependent and can be represented as
\begin{gather}
 Z_{2k}=\alpha^{2k-1}(kG+\alpha Z_0),\hskip 2cm
 Z_{2k-1}=\alpha^{2k-2}(kG-\alpha\bar Z_0),
 \notag\\\label{oad1}
 \bar Z_{2k}=\alpha^{2k-1}(kG+\alpha\bar Z_0),\hskip 2cm
 \bar Z_{2k-1}=\alpha^{2k-2}(kG-\alpha Z_0),
 \\\notag
 B_k=k\alpha^{k-1}B_1,
\end{gather}
where $k\in\N$. For the generators of this form the
mutually commuting operators (\ref{jmn}) are reduced to
\begin{equation}
 J_n^m=m\alpha^{m-1}J_n^1+\frac 14C_3^{m+1}\alpha^{m-3}I
 -\frac{(-1)^m+1}{16}m\alpha^{m-3}C^2,
\end{equation}
where $C^m_n=\frac{m!}{(m-n)!n!}$, $m\in\N$.
As a consequence, the one-dimensional system has only one
even independent integral of motion being its Hamiltonian.

In the case of the contracted Dolan-Grady relations
($\alpha=0$) the superpotential acquires the polynomial form
\begin{equation}\label{x2}
 W(x)=w_2 x^2 + w_1x+w_0,
\end{equation}
where all the parameters $w_0,w_1,w_2$ are real and,
as a consequence, the nonlinear holomorphic supersymmetry
turns out to be spontaneously broken \cite{d1}
(see, however, the next section).
In this
case the Onsager algebra does not appear since from
(\ref{oad1}) it follows that all the generators of $\OA$
vanish in the limit $\alpha\to 0$ except for $Z_0$, $\bar Z_
0$ and $B_1$. Similarly, except for $J_n^1$, all the
integrals (\ref{jmn}) vanish as well. Thus, from the point
of view of the Onsager algebra this contracted system is
trivial. One can verify that the resulting algebra obtained
from (\ref{ad1}) by the contraction $\alpha\to 0$ is a
nilpotent algebra.

\subsection{2D $n$-HSUSY}

Representation of the
nonlinear holomorphic supersymmetry on a plane
supplies us with a new, in comparison with
the 1D case, phenomenon of
the appearance of
nontrivial even central charges.
In terms
of complex coordinates $z=\frac 12(x_1+ix_2)$,
$\bar z=\frac 12(x_1-ix_2)$, the generating elements can be
represented as
\begin{align}\label{zd2}
 Z_0&=\partial+W(z,\bar z), &
 \bar Z_0&={}-\bar\partial+\bar W(z,\bar z).
\end{align}
In this representation the Hamiltonian (\ref{Halg})
corresponds to a charged spin-1/2 particle with
gyromagnetic ratio $g=2n$ moving in an external magnetic
field $B=\partial_1A_2-\partial_2A_1$ defined by the gauge
vector potential with components $A_1=\mathop{\sf Im}W$,
$A_2=\mathop{\sf Re}W$ \cite{d2}. The Dolan-Grady relations
give rise to the magnetic field of the form
\begin{equation}\label{bd2}
 B=w_+e^{\sqrt{2\alpha}x_1}+w_-e^{-\sqrt{2\alpha}x_1}
 +we^{i\sqrt{2\alpha}x_2}+\bar we^{-i\sqrt{2\alpha}x_2}
\end{equation}
with $w_\pm\in\R$, $w\in\C$, $\bar w=w^*$.
Unlike the one-dimensional case, the system
defined by the
representation (\ref{zd2}), (\ref{bd2})
admits in addition to the
Hamiltonian (\ref{Halg})
another independent even integral of motion being
a central charge
${\cal J}^2_n$ \cite{d2}:
\begin{equation}\label{j2}
 2{\cal J}^2_n=
 -\frac 14\left(\omega^2\bar Z^2+\bar\omega^2Z^2\right)
 +\partial B\bar Z+\bar\partial BZ - B^2
 +\frac n2\bar\partial\partial B\sigma_3,
\end{equation}
where $Z=Z_0$, $\bar Z=\bar Z_0$.

Let us discuss this system in the context of the Onsager
algebra. The generating elements in representation
(\ref{zd2}), (\ref{bd2}) define the finite-dimensional
algebra
\begin{align}
 [Z_0,\:\bar Z_0]&=B_1,&
 [Z_0,\:\bar G]&=D,&
 [G,\:\bar Z_0]&=D,\notag\\\label{ad2}
 [Z_0,\:B_1]&=G,&
 [Z_0,\:G]&=2\alpha B_1,&
 [\bar G,\:\bar Z_0]&=2\alpha B_1,\\\notag
 [B_1,\:\bar Z_0]&=\bar G,&
 [Z_0,\:D]&=2\alpha\bar G,&
 [D,\:\bar Z_0]&=2\alpha G,
\end{align}
where $\bar G=G^\dag$, $D^\dag=D$,
and $B_1$, $G$, $\bar G$ and $D$ commute between themselves. 
So, in the 2D case we
have the algebra with six independent generators.
This algebra has the two Casimir operators
\begin{align*}
 I_1&=B_1D-\bar GG,&
 I_2&=4\alpha^2B_1^2+D^2-2\alpha(G^2+\bar G^2).
\end{align*}
In representation (\ref{zd2}), (\ref{bd2}) they are
proportional to the unit operator,
$I_1=2^5\alpha(w_+w_--\bar ww)\cdot\I$,
$I_2=2^7\alpha^2(\bar ww+w_+w_-)\cdot \I$. Therefore,
the representation of the algebra (\ref{ad2})
in this case is
irreducible.
Then, rescaling the coordinates $x_1$, $x_2$,
one gets the set
$\{i\partial_{x_1},
i\partial_{x_2},e^{\pm x_1},
e^{\pm ix_2}\}$
as independent generators
forming the basis of the algebra
(\ref{ad2}). Therefore, this algebra is
$e(2)\oplus iso(1,1)$ with the Casimir operators to be
linear combinations of $I_1$ and $I_2$.

Similarly to the 1D case, the number of independent
continuous parameters of the two-dimensional system
corresponding to the 5-parametric magnetic field (\ref{bd2})
is equal to two:
using the rescaling and shift of the coordinates,
one can fix three parameters. The two independent
parameters characteristic
to the system with $n$-HSUSY, $w_+w_-$ and $\bar ww$, are
defined by linear combinations of the Casimir operators of
the algebra (\ref{ad2}) (or one can say that
they fix the values of the Casimir operators).

The Onsager algebra generated in the representation
(\ref{zd2}), (\ref{bd2}) is nontrivial. Like in the
one-dimensional case, all it's higher generators are
linearly
dependent but non-vanishing. In terms of the generators
of the algebra (\ref{ad2}) they are represented
in the form
\begin{align}
 Z_{2k}&=\alpha^{2k-1}(k\bar G+\alpha Z_0),&
 Z_{2k-1}&=\alpha^{2k-2}(kG-\alpha\bar Z_0),&
 B_{2k-1}&=(2k-1)\alpha^{2k-2}B_1,\nonumber\\
 \bar Z_{2k}&=\alpha^{2k-1}(kG+\alpha\bar Z_0),&
 \bar Z_{2k-1}&=\alpha^{2k-2}(k\bar G-\alpha Z_0),&
 B_{2k}&=k\alpha^{2k-2}D,\label{ad2g}
\end{align}
where $k\in\N$. Here the integrals (\ref{jmn})
are reduced to
\begin{align*}
 J_n^{2k+1}&=(2k+1)\alpha^{2k}J_n^1
 -\frac 14C_3^{2k+2}\alpha^{2k-2}I_1,&
 J_n^{2k}&=k\alpha^{2k-2}J_n^2
 -\frac 12C_3^{k+1}\alpha^{2k-4}I_2,
\end{align*}
i.e. all the commuting charges are linear functions of the
two independent integrals of motion (\ref{Halg}),
(\ref{j2}).

In the case of the contracted Dolan-Grady relations
($\alpha=0$) the magnetic field acquires the polynomial form
\cite{d2}
\begin{equation*}
 B=w_2\bar zz+\bar wz+w\bar z+w_0.
\end{equation*}
As a result, the corresponding Onsager algebra
turns out to be trivial since from the relations
(\ref{ad2g}) it follows that all the higher generators of
$\OA$ vanish in the limit $\alpha\to 0$ and the only
nontrivial generators are
$Z_0$,
$\bar Z_0$ and $B_1$. Similarly, except for $J_n^1$,
$J_n^2$,
all the integrals (\ref{jmn}), vanish. Like in the 1D case,
the resulting algebra obtained from (\ref{ad2}) by the
contraction $\alpha\to 0$ is a nilpotent algebra as well.

One notes that $n$-HSUSY
can also be realized in the 2D systems with nontrivial
Riemannian
geometry. 
This case is treated in detail in Ref.  
\cite{sphere}.

\section{Nonlinear pseudo-supersymmetry}

Till the moment we have discussed the Onsager algebra
and its realizations in the context of nonlinear holomorphic
supersymmetry with assumption of
the involution (\ref{invol}).
However, one can
discard (\ref{invol}) and regard the
generating elements of $\OA$ as Hermitian operators:
\begin{equation}
Z_0^\dagger=Z_0,\hskip 2cm
\bar Z_0{}^\dagger=\bar Z_0.
\label{zzd}
\end{equation}
In this
case the operators $B_m$ become anti-Hermitian, and,
as a consequence, the commuting charges (\ref{jmn})
are not
Hermitian operators any more.
On the other hand, one can introduce the operator
$\eta$ such that
\begin{equation}\label{ph}
\left({\cal J}^m_n\right)^\dag=\eta{\cal J}^m_n\eta^{-1}.
\end{equation}
In conventional matrix representation of the odd
operators,
\begin{equation}\label{theta}
\theta^\pm=\frac 12(\sigma_1\pm i\sigma_2),
\end{equation}
$\eta$ can be chosen, e.g., in the form
$\eta=\sigma_1$.
Then the relation (\ref{ph})
means that ${\cal J}^m_n$
can be treated as self-conjugate
operators with
respect to the indefinite scalar product
\begin{equation}
\label{scalpro}
(\psi_1|\psi_2)\equiv\langle\psi_1|\eta\psi_2\rangle=
\langle\psi_1\eta|\psi_2\rangle,
\end{equation}
where
$\langle\psi_1|\psi_2\rangle$
is the original positively definite scalar product.
In other words,
in the case of $\OA$
with generating elements
(\ref{zzd})
the operation of Hermitian conjugation
can be changed for the pseudo-Hermitian conjugation,
\begin{equation}
\label{pseh}
O^\ddag=\eta^{-1}O^\dag\eta,
\end{equation}
with respect to which the commuting charges
are invariant:
\begin{equation}
\label{psej}
{\cal J}^m_n{}^\ddag={\cal J}^m_n.
\end{equation}
Such a construction is very well known
and appears,
e.g., in the Dirac theory
for spin-1/2 particles.
Recently it emerged in the context of the
$PT$-invariant systems with real spectrum
\cite{znojil,phsym}. The property of the form
(\ref{ph})
named the $\eta$-pseudo-Hermiticity is
the necessary condition for having real spectra
in the systems with the Hamiltonian
satisfying the relation of the form (\ref{psej})
instead of to be a Hermitian operator
\cite{phsym}.
There, the
operator $\eta$ was referred
to as a Hermitian linear
automorphism.
In our case for the choice
$\eta=\sigma_1$, the
supercharges $Q_n$, $\bar Q_n$
are pseudo-Hermitian operators, and,
being nilpotent,
they turn out to be
similar in this sense to the Hermitian BRST and anti-BRST
charges
\cite{HT}.
Hence, when we treat the generating elements of
$\OA$ as Hermitian operators, some sort of
the nonlinear
supersymmetry arises as well.
Note, however, that unlike
the present case,
in the linear
pseudo-supersymmetry \cite{phsym}
including the $PT$-symmetric supersymmetry
\cite{znojil} as a particular case,
the nilpotent supercharges
are mutually conjugate with respect
to the indefinite scalar product (\ref{scalpro}).

It is worth noting that another possible
way to overcome the non-Hermiticity
of the commuting charges
(\ref{jmn}) for the Onsager algebra with
Hermitian generating elements
consists in performing the formal change $n\to in$.
However, with such a prescription the nonlinear
supersymmetric construction would be lost.

The examples of the systems with
nonlinear supersymmetry realized
in the form of pseudo-supersymmetry
can be obtained within the framework
of the spin chain models.
First, we note that with
taking into account the identification
(\ref{Zdef}),
the Hamiltonian of the
Transverse Ising Chain
given by the operator
$J_0$ from Eq. (\ref{lin})
is
constructed in terms of the
Hermitian operators \cite{onsager}
\begin{align}\label{ising}
 Z_0&=\sum_{i=1}^L\sigma_3(i)\sigma_3(i+1),&
 \bar Z_0&=\sum_{i=1}^L\sigma_1(i).
\end{align}
The system is defined on a periodic
chain of length $L$ with $\sigma_\alpha(i)$, $\alpha=1,2,3,$
being the Pauli matrices describing local spin on a site
with number $i$.
Since the
operators (\ref{ising}) satisfy the Dolan-Grady
relations, they can be used for the construction
of the system with nonlinear
holomorphic supersymmetry
given in terms of the Hamiltonian
$H_n={\cal J}^1_n$
(\ref{sj}) and supercharges (\ref{q0}),
where the Pauli matrices $\sigma_a$, $a=1,2,3,$
are supposed to be commuting with the
matrices $\sigma_\alpha(i)$.
Such a Hamiltonian
is $\eta$-pseudo-Hermitian
with $\eta=\sigma_1$.
The number of independent commuting charges
${\cal J}_n^m$ in such a supersymmetric
model is the same as for the linear set (\ref{lin})
of the integrals corresponding to the Transverse Ising Model
since they are based on the same representation of the
Onsager algebra.

In the same way one can construct the nonlinear
pseudo-supersymmetry proceeding from the
Hermitian generating elements
\begin{align}\label{z01}
 Z_0&=\sum_{i=1}^L\sigma_3(i)\sigma_3(i+1),&
 \bar Z_0&=\sum_{i=1}^L\sigma_1(i)\sigma_1(i+1).
 \end{align}
The operators (\ref{z01}) satisfy
the Dolan-Grady relations
and correspond to
the X-Y model \cite{dg}
like the operators (\ref{ising})
correspond to the Transverse Ising Model.

One more representation of the Dolan-Grady relations can be
found in $Z_N$ spin models \cite{gehlen}
generalizing the Ising Model. In this case the
Hermitian generating elements are
\begin{align}\label{Nchain}
 Z_0&=\sum_{i=1}^L\sum_{n=1}^{N-1}
 \frac{P_i^nP_{i+1}^{N-n}}{1-\omega^{-n}},&
 \bar Z_0&=\sum_{i=1}^L\sum_{n=1}^{N-1}
 \frac{X_i^n}{1-\omega^{-n}},
\end{align}
where $N=2, 3, \ldots$, and $X_i$, $P_i$ are
the local $Z_N$
spin
operators satisfying the relations
$[X_i,X_j]=[P_i,P_j]=0$,
$P_iX_j=\omega^{\delta_{ij}}X_jP_i$, $P_i^N=X_i^N=\I$,
$\omega=\exp(2\pi i/N)$. In the same line as in the
previous two cases, one can construct the system with
the pseudo-Hermitian
Hamiltonian possessing
the nonlinear supersymmetry.

It is worth noting that for all these spin models, one can
construct the generating elements of the second generation,
$Z_0^{(1)}$, $\bar Z_0^{(1)}$, and so on till some number
defined by the length $L$ of the periodic chain.

The described construction of the nonlinear
pseudo-supersymmetry with some modification
can be applied also to the quantum mechanical systems
discussed in the previous section.
As a result, one can obtain some quasi-exactly solvable
systems which appear in the context of the $PT$-symmetric
quantum mechanics.
Let us obtain one such a family of the 1D quantum mechanical
systems proceeding from the generators (\ref{zd1})
with the quadratic superpotential (\ref{x2}).
For such systems, unlike the $n$-supersymmetric
class of systems with exponential form of the
superpotential (\ref{wd1}),
the corresponding supersymmetry is always
spontaneously broken when
$w_2\ne 0$
\cite{d1}, while the case $w_2=0$ corresponds
to the superoscillator with
nonlinear supersymmetry to be the
exact symmetry of the system \cite{susy-pb}.
The zero modes of the supercharge
$Q_n$ with the superpotential (\ref{x2})
have the following leading factor for $|x|\to\infty$:
\begin{equation}\label{zero}
 \psi_0\sim e^{{}-\frac{w_2}3x^3-\frac{w_1}2x^2-w_0x}.
\end{equation}
Hence, such functions are not normalizable for $w_2\ne 0$.
Note, however,
that if the parameter $w_2$ is pure imaginary and
$w_1>0$, then the wave functions of the form (\ref{zero})
are
normalizable. But in this case the Hamiltonian (\ref{Halg})
is not Hermitian. If $w_0$ is also a pure imaginary number,
there exists an operator $\eta$ such that the Hamiltonian
(\ref{Halg}) is an $\eta$-pseudo-Hermitian operator. Indeed,
let us put
$\eta=P$, where $P$ is the parity (reflection)
operator, $PxP=-x$, $P^2=1$.
Then, with respect to the
pseudo-Hermitian
conjugation (\ref{pseh}), the operators (\ref{zd1})
are transformed
as ${Z^\ddag=-\bar Z}$, $\bar Z^\ddag=-Z$.
As a consequence, the Hamiltonian (\ref{Halg}) is
pseudo-Hermitian. Thus, in this case we have the
system with the nonlinear supersymmetry generated by the
supercharges (\ref{q0}) having the properties
$Q_n^\ddag=(-1)^n\bar Q_n$, $\bar Q_n^\ddag=(-1)^nQ_n$.
One notes
that although the parameters $w_0,w_2$
are pure imaginary, the anticommutator of the supercharges
is a
polynomial in the Hamiltonian
with real coefficients. This property
is in accordance with the generic property of any
$\eta$-pseudo-Hermitian system for which
the spectrum
consists of real and/or
complex conjugate pairs of numbers \cite{phsym}.

Rescaling the variable $x$,
the obtained system given by the
$\eta$-pseudo-Hermitian Hamiltonian
and possessing nonlinear supersymmetry
can be related to the three-parameter class of quasi-exactly
solvable systems with quartic polynomial potentials
discussed in Ref. \cite{bb} within the framework
of the $PT$-symmetric quantum mechanics.
One of the parameters taking there natural values,
in our construction corresponds to the parameter
$n$ defining the order of the nonlinear supersymmetry.

\section{Discussion and outlook}

Let us summarize briefly the obtained results and
discuss some open problems that deserve
further attention.

\begin{itemize}
 \item
 It was ascertained that the Onsager algebra associated with
 the Dolan-Grady relations underlies the
 nonlinear holomorphic  supersymmetry.
\end{itemize}
The Onsager algebra arisen in the framework of the
nonlinear holomorphic
supersymmetry is endowed with a natural
Hermitian conjugation. This involution is different from
that associated with the Onsager algebra usually discussed
in
the context of integrable models with the infinite set of
conserved charges constructed by Dolan and Grady. This is
the main peculiarity of our treatment of the Onsager
algebra.
\begin{itemize}
 \item
 The contracted Onsager algebra corresponding to the
 contracted Dolan-Grady relations was introduced.
\end{itemize}
As far as we know such an algebra was not discussed earlier.
This algebra has the properties similar to those of the
original
Onsager algebra but its structure is much more simple.
Besides, for some systems the Dolan-Grady relations can be
realized in the contracted form only. The examples of such
systems are given by
the
superoscillator with nonlinear superalgebra \cite{susy-pb},
the single-mode parafermions \cite{susy-pf},
a charged spin-1/2 particle
with gyromagnetic ratio $g=2n$ subjected to an external
magnetic field and moving in 2D space with non-trivial
Riemannian geometry \cite{sphere}.
\begin{itemize}
 \item The new infinite set of the commuting charges
 associated with the Onsager algebra was constructed.
\end{itemize}
Unlike the set (\ref{lin}) found by Dolan and Grady, in the
new set (\ref{jmn}) the charges have the
terms quadratic in the
generators of the Onsager algebra. The operators contain an
arbitrary real parameter that can be interpreted as a
coupling constant. We have shown that $n$-HSUSY
is associated with this system
for the coupling constant to be integer. In other words, for
$n\in\Z$ there exits an intertwining operator that relates
the spectra of the operators $J^m_n$ and $J^m_{-n}$ for any
$m\in\N$.

\begin{itemize}
 \item
 The notion of the nonlinear holomorphic supersymmetry was
 extended to the case of nonlinear pseudo-supersymmetry.
\end{itemize}
It seems that this construction could be useful for
investigation of different $PT$-symmetric and other systems
with pseudo-Hermitian Hamiltonians, the
interest to which has
considerably grown recently \cite{phsym,znojil}.

In the context of the Onsager algebra construction,
we clarified also the nature of the known 1D and 2D
systems with nonlinear holomorphic
supersymmetry \cite{d1,d2}.

It is worth noting that the Onsager algebra admits
a reformulation as a Poisson algebra of some mechanical
system (see Ref. \cite{kz} for some
classical realization of
the Dolan-Grady relations).
A priory, such a corresponding
system may also possess the infinite set of
commuting integrals, which could be obtained from
(\ref{jmn})
by the formal change $\{\bar Z,Z\}\to 2\bar Z Z$.
Besides,
the contracted Onsager algebra can be represented as a
Poisson algebra of a continuous symplectic system. It would
be interesting to find an explicit nontrivial example of
such a classical system. If such hypothetical
system realizes the
infinite dimensional contracted Onsager algebra
with linearly independent generators,
then this system would be
integrable due to existence of the
infinite set of commuting integrals of motion. We hope that
further investigation will shed light on this problem.

\vskip 0.5cm
{\bf Acknowledgements}
\vskip 5mm

The work was supported by the grants 1010073
and 3000006 from FONDECYT (Chile) and by DYCIT
(USACH). M.P. thanks A. Zhedanov for
a very useful communication.

\end{document}